\documentstyle[11pt]{article}

\title{Computation in an algebra of test selection criteria}

\author{Jan Pachl\thanks{IBM Research Division,
                         Zurich Research Laboratory,
                         S\"{a}umerstrasse 4,
                         8803 R\"{u}schlikon,
                         Switzerland}
        \ and Shmuel Zaks\thanks{Department of Computer Science,
                         Technion,
                         Haifa,
                         Israel}
       }
\date{May 12, 1993}

\newcommand{\siz}[1]{\mbox{$|#1|$}}
\newcommand{\probl}[3]{\vspace{2mm}\noindent {\bf #1} \\
                       \indent{\bf Input:} #2    \\
                       \indent{\bf Question:} #3 \\
                       \vspace{-5mm}\mbox{}}
\newcommand{\sprobl}[3]{\vspace{2mm}\noindent {\bf #1} \\
                       \indent{\bf Input:} #2    \\
                       \indent{\bf Output:} #3 \\
                       \vspace{-5mm}\mbox{}}
\newcommand{\fromto}[4]{\parbox{#1}{
                \begin{center}
                \rule{0cm}{3mm}${\scriptstyle #4}$ \\[-3mm]
                \rule{0cm}{4mm}$#2$                \\[-3mm]
                \rule{0cm}{3mm}${\scriptstyle #3}$
                \end{center}       }
                       }
\newcommand{\MIN}{\mbox{${\sf MIN}$}}

\newcommand{\MumAS}{\mbox{${\sf MumAS}$}}
\newcommand{\MalAS}{\mbox{${\sf MalAS}$}}
\newcommand{\EA}{\mbox{${\sf EA}$}}
\newcommand{\MASI}{\mbox{${\sf MASI}$}}
\newcommand{\uce}{\mbox{$\union\cross{\scriptstyle =}$}}
\newcommand{\iiiSAT}{\mbox{${\sf 3SAT}$}}
\newcommand{\GRAPHCOL}{\mbox{{\sf GRAPH}} \mbox{{\sf K-COLORABILITY}}}
\newcommand{\exh}{\mbox{${\sf EXHAUSTIVE}$}}
\newcommand{\ea}{\mbox{${\sf EACH}$}}

\newcommand{\critone}{\mbox{${\sf ANY\_TEST}$}}
\newcommand{\pil}{\mbox{${\sf S}$}}

\newcommand{\union}{\mbox{$ \uplus $}}
\newcommand{\cross}{\mbox{$ \otimes $}}
\newcommand{\bigunion}{\mbox{$ \biguplus $}}
\newcommand{\bigcross}{\mbox{$ \bigotimes $}}
\newcommand{\unionand}{\mbox{$ \wedge\uplus $}}
\newcommand{\crossand}{\mbox{$ \wedge\otimes $}}
\newcommand{\unionor}{\mbox{$ \vee\uplus $}}
\newcommand{\crossor}{\mbox{$ \vee\otimes $}}
\newcommand{\qvx}{\mbox{$ ( q_1 , \ldots , q_N ) $}}
\newcommand{\vvx}{\mbox{$ ( v_1 , \ldots , v_N ) $}}

\newcommand{\param}{\mbox{$ \Delta $}}
\newcommand{\constr}{\mbox{$ \psi $}}
\newcommand{\criter}{\mbox{$ \Gamma $}}
\newcommand{\seq}{\mbox{$ = $}}    % strong equality of criteria
\newcommand{\weq}{\mbox{$ \:\simeq\: $}} % weak equality of criteria
\newcommand{\wleq}{\mbox{$ \:\sqsubseteq\: $}} % weak comparison

\newcommand{\eqmerge}{\;\|\;}
\newcommand{\sepa}{\mbox{\bf separator\_1}}
\newcommand{\sepb}{\mbox{\bf separator\_2}}
\newcommand{\stra}{\mbox{\bf string\_1}}
\newcommand{\strb}{\mbox{\bf string\_2}}
\newcommand{\aoccurs}{\mbox{\bf string\_1\_occurs}}
\newcommand{\vslash}{\mbox{{"{\tt /}"}}}
\newcommand{\vzet}{\mbox{{"{\tt z}"}}}
\newcommand{\veks}{\mbox{{"{\tt x}"}}}
\newcommand{\vempty}{\mbox{{""}}}
\newcommand{\va}{\mbox{{"{\tt a}"}}}
\newcommand{\vab}{\mbox{{"{\tt ab}"}}}
\newcommand{\vabcd}{\mbox{{"{\tt abcd}"}}}
\newcommand{\vabcdnine}{\mbox{{"{\tt abcd987}"}}}
\newcommand{\vlong}{\mbox{{"{\tt abcdefghijklmnopqrstuvwxyz0123}"}}}
\newcommand{\et}{\mbox{\rm\ \  and \ \ }}
\newcommand{\true}{\mbox{\sf true}}
\newcommand{\false}{\mbox{\sf false}}
\newcommand{\qed}{ \hspace*{\fill} \hspace*{0.5in} $\Box$
                   \hspace*{0.5in} \linebreak \mbox{} }

\newcommand{\fuline}{\rule{15cm}{0.3mm}}

\newcommand{\ignore}[1]{}

\newtheorem{theorem}{Theorem}[section]
\newtheorem{lemma}[theorem]{Lemma}
\newtheorem{proposition}[theorem]{Proposition}
\newtheorem{corollary}[theorem]{Corollary}

\begin{document}

\maketitle

\vspace{5mm}

\begin{abstract}
One of the key concepts in testing is that of adequate test sets.
A {\em test selection criterion\/} decides which test sets are adequate.
In this paper, a language schema for specifying a large class of test
selection criteria is developed;
the schema is based on two operations for building complex criteria
from simple ones.
Basic algebraic properties of the two operations are derived.

In the second part of the paper, a simple language --- an instance of
the general schema --- is studied in detail, with the goal of generating
small adequate test sets automatically.
It is shown that one version of the problem is intractable,
while another is solvable by an efficient algorithm.
An implementation of the algorithm is described.
\end{abstract}

\vspace{2cm}

Note added on December 23, 1999 \{Jan Pachl\}: This version is
dated May 12, 1993. A previous version was issued as an IBM
research report RZ 2114, dated April 16, 1991. The paper has not
been published.

\clearpage
\tableofcontents

\clearpage
\section{Introduction}
      \label{sec:intro}

This paper deals with testing of computer programs.
However, most of our discussion applies to testing of more general
systems.

Testing consists of experiments, called {\em tests\/},
in which the behavior of
the system under test is compared to its {\bf specification}.
The system is often called an {\bf implementation under test};
the purpose of testing is to conclude whether the system
implements the specification.

The test designer must decide, possibly with machine assistance,
what tests are to be executed and in what order.
In this paper we assume that tests are repeatable and that the behavior
of the implementation under test in each individual test does not depend
on the order in which the tests are executed.
Therefore the test designer's decision is described by a {\em set of tests\/},
selected from some set of tests that {\em could\/} be executed.
To model this situation, we denote by $D$ the {\bf test domain\/},
i.e. some given set of tests for the implementation under test.
Subsets of $D$ are called {\bf test sets\/}.

An important concept is that of adequate test sets.
Informally, a subset $T$ of $D$ is {\bf \-adequate\/} if we believe that
it is sufficient to execute the tests in $T$,
instead of all the tests in $D$.
Once we have checked that
the behavior of the implementation satisfies the specification
for each test $d$ in $T$,
we are willing to accept that the same will be true for each $d$ in $D$.
To make this concept independent of subjective beliefs,
we define adequacy with respect to a test selection criterion:
A {\bf test selection criterion on\/} $D$ is a rule that decides
for each subset $T$ of $D$ whether $T$ is adequate or not.
(Other terms have been used in the literature,
e.g. {\em data selection criterion\/}~\cite{A0894},
{\em test method\/}~\cite{gourlay},
{\em testing method\/}~\cite{hamlet}).
A test selection criterion may be defined based on the knowledge of
the implementation under test, of its specification, or both;
Gourlay~\cite{gourlay} introduced a framework for discussing these
dependencies explicitly.

Many natural test selection criteria can be described as follows:
There is a collection of subsets of the domain $D$, and $T\subseteq D$
is adequate if and only if $T$ intersects every nonempty set
in the collection.
The following three examples of selection criteria from
the literature, and many others, are of this form.
\begin{enumerate}
    \item
       {\bf Condition table method}~\cite{A0894}.
       ``[I]dentify conditions describing some aspect of the problem
       or program to be tested''~(\cite{A0894}, p.~167),
       and then combine the conditions to form test predicates on~$D$,
       the set of inputs.
       A test set $T$ is {\em complete\/}~(\cite{A0894}, p.~170) if
       \begin{itemize}
           \item
              for each thus formed test predicate there is a point in $T$
              that satisfies the predicate; and
           \item
              each point in $T$ satisfies at least one of the predicates.
       \end{itemize}
       The first condition is clearly the adequacy of $T$ as described above,
       with respect to a collection of subsets of $D$.
    \item
       {\bf Cause-effect graphing}~\cite{ceg,myers}.
       A {\em cause-effect graph\/} is a simplified specification
       of the system under test.
       Nodes in the graph represent important properties of {\em causes\/}
       (inputs) and {\em effects\/} (outputs) and possibly additional
       intermediate properties.
       Edges represent how the effects depend on the causes.
       Once the cause-effect graph has been constructed, it can be used
       for systematic selection of a set of inputs for testing.
       Let~$N$ be the set of nodes in the graph.
       Each input defines a subset of~$N$;
       thus the domain~$D$ corresponds to a set of subsets of $N$.
       One simple test selection criterion is:
       \begin{itemize}
           \item
              Ensure that each effect node is covered at least once.
       \end{itemize}
       This is clearly adequacy as described above,
       with respect to a collection of subsets of~$D$.
       Myers~(\cite{myers}, pp.~65-68) described a more complex
       test selection criterion based on the cause-effect graph;
       again his description can be defined as adequacy with respect
       to a collection of subsets of $D$.
    \item
       {\bf Statement coverage}~\cite{myers}.
       Let the implementation under test be implemented by
       a program consisting of a number of statements.
       For each statement $s$ in the program, let $X_s$ be the set of
       the tests in $D$ that cause $s$ to be executed.
       Then $T \subseteq D$ is adequate with respect to the collection
       $ \{ X_s \} $
       if and only if $T$ covers every statement covered by $D$.
\end{enumerate}
Jeng and Weyuker~\cite{jeng} give several other examples of test selection
criteria of this general form, which they call {\em partition testing\/}.

In the present paper we describe a simple but powerful language
for specifying test selection criteria;
the language is based on our previous proposal~\cite{notation}.
A language for specifying test selection criteria is needed when we wish
to free the test designer from dealing with individual test cases.
The test designer should be able to specify
what constitutes an adequate test set in a high-level notation,
from which individual test cases are then generated automatically.

Balcer, Hasling and Ostrand~\cite{balcer} built a system called TSL,
which supports this high-level approach to testing.
Our design can serve as a model for extending the test specification
language in TSL, and for defining other similar languages.
We return to the comparison with TSL in Section~\ref{subsec:compareTSL}.

We describe a general language {\em schema\/},
from which concrete languages are derived by choosing types of parameters.
The schema is based on two operations for combining selection criteria;
with these two operations, test selection criteria form a well-behaved
algebra.
The ability to combine criteria using the two operations yields
a number of benefits:
\begin{itemize}
    \item
       The language has simple well-defined semantics.
    \item
       The language is powerful --- many useful criteria can be expressed
       in the language.
    \item
       Algorithms that process criteria and generate test sets can use
       algebraic identities to manipulate criteria.
\end{itemize}

In the second half of the paper we define one language based on the general
schema, and study the algorithms that generate adequate test sets for
the criteria expressed in the language.
We show that the problem of finding a minimum adequate test set
(i.e. an adequate test set of the smallest size) is NP-hard,
and then we concentrate on the problem of finding a minimal adequate
test set (i.e. a test set whose proper subsets are not adequate).
We also describe what we learned from implementing a prototype tool
for generating minimal adequate test sets.

Related work
and topics for further research are discussed in the last section.

\section{Example}
    \label{sec:example}

To illustrate the concept of a test selection criterion,
we now describe a simple testing scenario,
adopted from the paper by Balcer, Hasling and Ostrand~\cite{balcer}.

Test suites typically consist of many test cases that differ
only slightly from each other.
Rather than preparing all the variations one by one,
the test designer may prepare
a ``parameterized test case''
(a ``code template'' in the terminology of~\cite{balcer})
and then generate individual
test cases by systematically filling in the values of the parameters.

In the sample scenario, a text editor is to be tested against the specification
of the CHANGE command.
The syntax of the command is \\
\hspace*{2cm} {\tt C /string1/string2\/}   \\
As in~\cite{balcer}, the parameterized test case for this task uses
five parameters.
(More precisely, the TSL description in~\cite{balcer} uses four
parameters and one environment condition;
however, the distinction is not important for our discussion.)

Parameter declarations are in Figure~\ref{fig:decl}.
To obtain one individual test case, we select one value
for each parameter, and substitute the selected values to \\
\hspace*{2cm} {\tt C\/} \ \sepa\ \ \stra\ \ \sepb\ \ \strb  \\
The value of the parameter \aoccurs\ is used to set up the current line
in the editor (so that it does or does not contain \stra).

\begin{figure}[tb]

\fuline
\begin{tabbing}
\hspace{3mm}\=\hspace{8mm}\=\hspace{8mm}\=\hspace{8mm}\=\hspace{8mm}\=
                                                                \kill
{\sf declaration}
\\[-2.5mm] \>\sepa  \>\>\>: \{ \vslash, \vzet \,\}
\\[-2.5mm] \>\sepb  \>\>\>: \{ \vslash, \veks \,\}
\\[-2.5mm] \>\stra  \>\>: \{ \vempty, \va, \vab, \vabcd,
\vabcdnine, \vlong \,\}
                                                                \\[-2.5mm]
\>\strb  \>\>: \{ \vempty, \va, \vab, \vabcd, \vabcdnine, \vlong \,\}
                                                                \\[-2.5mm]
\>\aoccurs  \>\>\>\>: \{ \true, \false \,\}
\end{tabbing}
\fuline
\caption{Parameter declarations for the example}
\label{fig:decl}
\end{figure}

Now observe that the parameter declarations in Figure~\ref{fig:decl}
define a test domain $D$:
Each combination of values for the five parameters defines a test in $D$.
In some cases it may be feasible to execute all tests in $D$.
However, even in our simple example $D$
contains $ 2 \times 2 \times 6 \times 6 \times 2 = 288 $ elements.
It is easy to imagine much larger examples, for which testing
with all inputs in $D$ would be infeasible.
The test designer must then select a {\bf test set\/},
i.e. a subset of $D$.
Sometimes the test designer wants to list the points of the test
set explicitly, one by one.
However, it is frequently more convenient
to write a high-level description of a test selection criterion,
and let an automated tool select a test set adequate for the criterion.

Let us consider several examples of high-level descriptions
of test selection criteria that free the test designer
from the need to think in terms of individual test cases.
For our domain $D$, the criterion
\begin{equation}
    \langle\; \mbox\stra = \va \;\rangle
\end{equation}
specifies that the test set must include at least one point
in which the value of the parameter \stra\ is \va.
The criterion
\begin{equation}
     \ea ( \;\stra \; : \;\va, \;\vab, \;\vabcdnine \;)
                                     \label{eq:eaA}
\end{equation}
specifies that for each of the three listed values
of the parameter \stra\
the test set must include at least one point with that value.
It is convenient to have another primitive as an abbreviation
for \ea\ whose arguments include all values declared for
the parameter; the primitive \exh\ with one argument has this role.
Thus
\begin{equation}
    \exh( \;\stra \;)
                          \label{eq:exhA}
\end{equation}
has the same meaning as
\begin{equation}
    \ea ( \;\stra \; :
          \;\vempty, \;\va, \;\vab, \;\vabcd, \;\vabcdnine, \;\vlong \;)
          \;\; .
                                     \label{eq:eaAa}
\end{equation}

As we shall see in the next section, (\ref{eq:exhA})
and the criterion
\begin{equation}
    \exh( \;\sepa \;)
                          \label{eq:exhSA}
\end{equation}
can be combined in two basic ways.
One combination is
\[
    \exh( \;\stra \;)  \;\;\cross\;\;  \exh( \;\sepa \;) \;\; ,
\]
which can be also written as
\[
    \exh( \;\stra , \;\sepa \;) \;\; .
\]
It specifies that all possible combinations of the values
of \stra\ and \sepa\ must be included;
since \stra\ assumes six values and \sepa\ two values,
any test set adequate for this criterion must contain at least
12 elements.
The other combination of~(\ref{eq:exhA}) and~(\ref{eq:exhSA}) is
\[
    \exh( \;\stra \;)  \;\;\union\;\;  \exh( \;\sepa \;) \;\; ,
\]
which merely requires that the test set must be adequate
for~(\ref{eq:exhA}) and also for~(\ref{eq:exhSA}).
A test set containing 6 points is sufficient for that;
for example, the following six combinations of
\stra\ and \sepa\ are sufficient:
\begin{tabbing}
\hspace{4cm}\=\hspace{1cm}\=\hspace{7cm}\=            \kill
\>\>\stra\>\sepa                                      \\[1.5mm]
\>1.\>\vempty\>\vslash                                \\[-2mm]
\>2.\>\va\>\vslash                                    \\[-2mm]
\>3.\>\vab\>\vslash                                   \\[-2mm]
\>4.\>\vabcd\>\vslash                                 \\[-2mm]
\>5.\>\vabcdnine\>\vslash                             \\[-2mm]
\>6.\>\vlong\>\vzet                                   \\[-2mm]
\end{tabbing}

In the next section we describe a more systematic approach to
the construction of test selection criteria.
We shall see that many complex criteria, including \ea\ and \exh,
may be constructed from simple ones.

\section{A language for test selection criteria}
    \label{sec:nota}

\subsection{A general language schema}
      \label{subsec:schema}

We are now going to describe a language for specifying instances
of the test selection problem.
We start by describing a general {\bf language schema\/}.
Many different concrete languages may then be obtained
from the schema by allowing different parameter types.
One such choice of parameter types and the resulting
concrete language are discussed
in Section~\ref{subsec:enum} and in the rest of the paper.

To define an instance of the test selection problem,
we have to specify
a domain $D$ and a test selection criterion on $D$.
In our approach,
$D$ and the criterion on $D$ have the following special form:
\begin{itemize}
   \item
   $D$ is a subset of the Cartesian product
   $ P = \displaystyle\prod_{i=1}^N Q_i $
   of certain sets $Q_1 , \ldots, Q_N$.
   The points in $P$ are vectors \vvx\
   of parameter values $ v_i \in Q_i $.
   \item
   The criterion is defined by a set of subsets of $P$.
\end{itemize}

Thus to define an instance of the test selection problem,
we specify sets $Q_i\,$, a subset $D$ of the product $P$ of $Q_i\,$,
and a set of subsets of $P$.
In our language, the specification consists of three parts:
\begin{enumerate}
   \item
   declaration of parameters;
   \item
   a constraint;
   \item
   a test selection criterion.
\end{enumerate}
Part~1 defines the sets~$Q_i\,$, part~2 the set~$D$, and part~3
the set of subsets of~$P$.

The first part, denoted \param,
is a set of {\bf declarations\/}
\[
    q_i \;\; : \;\; Q_i
\]
each of which declares a {\bf parameter\/} $q_i$
and its {\bf range\/} $Q_i$.
Define
\[
    P(\param) = \prod_{i=1}^N Q_i \;\; .
\]

For example, for the declarations in Figure~\ref{fig:decl},
$ P(\param) $ is the Cartesian product of five sets~$Q_i$:
\begin{eqnarray*}
Q_1 & = & \{ \vslash, \vzet \}          \\
Q_2 & = & \{ \vslash, \veks \}          \\
Q_3 = Q_4 & = & \{ \vempty, \va, \vab, \vabcd, \vabcdnine, \vlong \}   \\
Q_5 & = & \{ \true, \false \}
\end{eqnarray*}

The second part is a {\bf constraint\/}; it is a boolean expression
$\constr = \constr\qvx$ built from {\bf primitive constraints}
by means of binary operators
$\vee$ (logical or) and $\wedge$ (logical and).
To interpret the constraint, we have to assign the value
\true\ or \false\ to each primitive constraint in the expression
when arbitrary values \vvx\ are substituted for the parameters \qvx.
The constraint then defines the domain
\[
    D(\param,\constr) =
       \{\;\vvx \in P(\param) \;\;|\;\; \constr\vvx = \true\;\}\;.
\]
We write $D(\constr)$ instead of $D(\param,\constr)$ when no
misunderstanding is possible.

The example in Section~\ref{sec:example} does not specify any constraint,
and therefore $D(\constr) = P(\param) $.

The third part is a {\bf test selection criterion\/};
it is an expression built from {\bf primitive criteria}
by means of binary operators \union\ and \cross.
The value of such an expression \criter\ is
a set $\pil(\param,\criter)$ of subsets of $P(\param)$.
Again we write $\pil(\criter)$ instead of
$\pil(\param,\criter)$ when no misunderstanding is possible.
Once the value $\pil(\criter)$ has been defined
for every primitive criterion~\criter,
we define $\pil(\criter )$ for general \criter\ as follows:
Given two criteria $\criter_1$ and $\criter_2$, define
\begin{eqnarray*}
\pil( \criter_1 \;\union\; \criter_2 )  & = &
\{ \; X \;\; | \;\; X \in \pil( \criter_1 )
               \;\;\; \mbox{\rm or} \;\;\;
                X \in \pil( \criter_2 )\; \}
\;\; = \;\; \pil( \criter_1 ) \cup \pil( \criter_2 )
\;\; ,  \\
\pil( \criter_1 \;\cross\; \criter_2 )  & = &
\{ \; X_1 \cap X_2 \;\; |
\;\; X_1 \in \pil( \criter_1 ) ,
X_2 \in \pil( \criter_2 ) \; \} \;\; .
\end{eqnarray*}

In our example in Section~\ref{sec:example},
when \criter\ is the primitive criterion
\[
    \langle\; \mbox\stra = \va \;\rangle
\]
the set $\pil(\criter)$ contains a single subset of $P(\param)$,
namely
\[
   \{ \; ( sep_1 , sep_2 , s_1 , s_2 , o ) \in P(\param)
   \; | \; s_1 = \va \; \}  \;\; .
\]
Similarly, we could take \ea\ and \exh\ as primitive criteria
and define their values $\pil(\criter)$;
however, we shall see later that these criteria can be derived
from simpler ones using \union\ and \cross.

{\bf Definition.}
An {\bf instance of the test selection problem\/} is
\(
    I = (\; \param, \; \constr , \; \criter \;)
\),
where \param\ is a set of parameter declarations,
\constr\ is a constraint,
and \criter\ is a test selection criterion.
A set $T \subseteq D(\param,\constr)$ is {\bf adequate for} $I$ if
$ \; T \cap X \neq \emptyset \; $
for every $ X \in \pil(\param,\criter )$
such that $ X \cap D(\param,\constr) \neq \emptyset $.
We also say that $T$ is {\bf adequate for\/} \criter\,
if \param\ and \constr\ are understood from the context.

From the definition of $\criter_1 \union \criter_2 $
it follows that a test set $T$ is adequate for
$ \criter_1 \union \criter_2 $ if and only if it is adequate for
$ \criter_1 $ and also for $ \criter_2 $.
The criterion $ \criter_1 \union \criter_2 $
is used when the test designer
wants to satisfy $\criter_1$ and $\criter_2$ independently.

The criterion $ \criter_1 \cross \criter_2 $ is used when
the test designer suspects dependencies between $\criter_1$
and $\criter_2$,
and wants to test for the faults produced
by combinations of causes.
If $\criter_1$ enforces the selection of a test point that has
some property $p_1$ and
$\criter_2$ the selection of a test point that has some property $p_2$, then
the criterion $\criter_1 \cross \criter_2$ enforces the selection
of a test point with the property $p_1 \; \mbox{\bf and} \; p_2$
(if such a point exists in~$D(\constr)$).

Since $\pil(\criter)$ is the value of the expression \criter,
it is natural to write
$\criter_1 = \criter_2$ when $\pil(\criter_1) = \pil(\criter_2)$,
and $\criter_1 \subseteq \criter_2$ when
$\pil(\criter_1) \subseteq \pil(\criter_2)$.
It is a simple exercise to show that
both $\union$ and $\cross$ are commutative and associative,
and that the following distributive law holds:
\[
( \criter_1 \;\union\; \criter_2 ) \;\cross\; \criter_3
\; \seq \; ( \criter_1 \;\cross\; \criter_3 ) \;\union\;
( \criter_2 \;\cross\; \criter_3 )  \;\; .
\]

Since \union\ and \cross\ are associative,
we write expressions like
$\criter_1 \union \criter_2 \union \criter_3$ and
$\criter_1 \cross \criter_2 \cross \criter_3$ without
parentheses. We also use the notation
$ \fromto{6mm}{\bigunion}{j=1}{m}{\criter_j} $
for $\criter_1 \union \criter_2 \union \ldots \union \criter_m $,
and similarly for~\bigcross.

\subsection{Comparing criteria}

In this section we define several relations for comparing
test selection criteria.
The definitions of this section are not used in the rest of
the paper, but the concepts will illustrate some important
properties of the algebra of test selection criteria.

The following relation $\wleq$ describes the notion
that one criterion is less stringent than another.

{\bf Definition.}
Let $\pil_1$ and $\pil_2$ be two sets of subsets of a set $P$.
Write $\pil_1 \wleq \pil_2$ if the following is true
for every $T \subseteq P$:
if $T \cap X \neq \emptyset $ for every nonempty $ X \in \pil_2 $
then $T \cap X\neq \emptyset $ for every nonempty $ X \in \pil_1 $.
Write $\pil_1 \weq \pil_2$ if $\pil_1 \wleq \pil_2$
and $\pil_2 \wleq \pil_1$.
For a fixed \param\ and criteria $\criter_1$ and $\criter_2$,
write $\criter_1 \wleq \criter_2$ if
$\pil(\param,\criter_1 ) \wleq \pil(\param,\criter_2 )$,
and  $\criter_1 \weq \criter_2$ if
$\pil(\param,\criter_1 ) \weq \pil(\param,\criter_2 )$.

The proof of the following proposition follows directly from
definitions.
In view of part~1,
$\criter_1 \wleq \criter_2$ if and only if
$( \param, \true, \criter_2 )$
subsumes $( \param, \true, \criter_1 )$
in the terminology of Hamlet~\cite{hamlet}.

\begin{proposition}
   Let \param\ be a fixed set of declarations.
   If $\criter_1$ and $\criter_2$ are two criteria then
   \begin{enumerate}
   \item
      $\criter_1 \wleq \criter_2$ if and only if
      every $ T \subseteq P(\param) $ adequate for
      $( \param, \true, \criter_2 )$
      is also adequate for
      $( \param, \true, \criter_1 )$;
   \item
      $\criter_1 \weq \criter_2$ if and only if
      $( \param, \true, \criter_1 )$ and
      $( \param, \true, \criter_2 )$ have the same adequate sets;
   \item
      $ \criter_1 \subseteq \criter_2 $ implies
      $ \criter_1 \wleq \criter_2 $.
      \qed
   \end{enumerate}
\end{proposition}

By part~3,
$ \criter_1 = \criter_2 $ implies $ \criter_1 \weq \criter_2 $.
Although $ \criter_1 \weq \criter_2 $ does not imply
$ \criter_1 = \criter_2 $,
Proposition~\ref{prop:minpile} below shows that
$\weq$ and $=$ are closely related.

Let \pil\ be a set of subsets of a set $P$.
A set $ X \in \pil$ is {\bf minimal in \pil\/} if
$ X \neq \emptyset $ and
\[
   Y \in \pil , \;  Y \subseteq X
   \;\; \mbox{\rm implies} \;\; Y = X
   \;\; \mbox{\rm or} \;\; Y = \emptyset \;\;.
\]
Let \MIN(\pil) be the set of all minimal $ X \in \pil $.

\begin{proposition}
     \label{prop:minpile}
If \pil, $\pil_1$ and $\pil_2$ are finite sets of subsets of $P$ then
\begin{enumerate}
   \item
     $ \MIN(\pil) \weq \pil \; $ ;
   \item
     $ \pil_1 \weq \pil_2 $ if and only if
     $ \MIN(\pil_1 ) = \MIN(\pil_2 ) \; $ .
\end{enumerate}
\end{proposition}

{\em Proof.}
1. Since $ \MIN(\pil) \subseteq \pil $, it follows that
$ \MIN(\pil) \wleq \pil $.
Since \pil\ is finite, for every nonempty $ Y \in \pil$
there exists a minimal $ X \in \pil $ such that $ X \subseteq Y $;
therefore $ \pil \wleq \MIN(\pil) $ by the definition of $ \wleq $.

2. If $ \MIN(\pil_1 ) = \MIN(\pil_2 ) $ then by part~1 we get
\[
   \pil_1 \weq \MIN(\pil_1 ) = \MIN(\pil_2 ) \weq \pil_2 \; \; .
\]
Assume $ \pil_1 \weq \pil_2 $ and $ X_1 \in \MIN(\pil_1 ) $.
We have $ \MIN(\pil_1) \weq \MIN(\pil_2) $ by part 1.
Set $ T_1 = P \setminus X_1 $; thus $ T_1 \cap X_1 = \emptyset $.
By the definition of $ \MIN(\pil_1) \wleq \MIN(\pil_2) $
there exists $ X_2 \in \MIN(\pil_2) $ such that
$ T_1 \cap X_2 = \emptyset $.
Thus $ X_2 \subseteq X_1 $.
Now by the same argument applied to $ T_2 = P \setminus X_2 $
there exists $ X'_1 \in \MIN(\pil_1) $ such that
$ X'_1 \subseteq X_2 \subseteq X_1 $.
Since $X_1$ is minimal in $ \MIN(\pil_1) $,
we have $X_1 = X'_1$, and therefore $X_1 = X_2$.
We have proved that every $ X_1 \in \MIN(\pil_1 )$ belongs
to $ \MIN(\pil_2 ) $.
By symmetry we get $ \MIN(\pil_1 ) = \MIN(\pil_2 ) $.
\qed

It is easy to verify that
\begin{eqnarray*}
\criter_1 \;\wleq\; \criter'_1 \;\;\; & \mbox{\rm implies} & \;\;\;
\criter_1 \;\union\; \criter_2 \;\wleq\; \criter'_1 \;\union\; \criter_2
     \\
\criter_1 \;\weq\; \criter'_1 \;\;\; & \mbox{\rm implies} & \;\;\;
\criter_1 \;\union\; \criter_2 \;\weq\; \criter'_1 \;\union\; \criter_2
\end{eqnarray*}

However, $ \criter_1 \weq \criter'_1 $ does {\,\em not\/\,} imply
$ \criter_1 \cross \criter_2 \weq \criter'_1 \cross \criter_2 $.
Thus, even for $\constr = \true $,
to determine which test sets are adequate with respect to
$( \param, \constr, \criter_1 \cross \criter_2 )$,
it is not enough to know which test sets are adequate with respect to
$( \param, \constr, \criter_1 )$ and which are adequate with respect to
$( \param, \constr, \criter_2 )$.

\subsection{Enumerated types}
    \label{subsec:enum}

From the general language schema described in
Section~\ref{subsec:schema} we obtain a concrete language
by specifying allowed parameter types.
To specify a parameter type, we must describe
\begin{itemize}
   \item
   the range;
   \item
   primitive constraints;
   \item
   primitive criteria.
\end{itemize}
In addition, we must supply rules to evaluate primitive
constraints and primitive criteria, so that
$D(\constr)$ and $\pil(\criter)$ are defined for any
\constr\ and \criter.

We use the following convention:
If $\varphi = \varphi\qvx$ is a Boolean expression then
$ \langle \varphi \rangle $ is the criterion for which
the value $\pil(\langle \varphi \rangle )$ contains a single
subset of $P(\param)$, namely
\[
   \{ \; \vvx \in P(\param) \;\; | \;\; \varphi\vvx \; \} \; .
\]

In the rest of the paper we work with one concrete language
obtained as follows:
Each parameter range is a finite set,
which is explicitly listed in the declaration.
Each primitive constraint has one of the two forms
\begin{eqnarray*}
    q_i & = & c_i    \\
    q_i & \neq & c_i
\end{eqnarray*}
where $q_i$ is one of the declared parameters,
and $c_i$ is one of the values in the range of $q_i$;
it is obvious how these constraints evaluate to \true\ or \false.
Each primitive criterion has one of the three forms
\begin{eqnarray*}
   &  \langle \; q_i \; = \; c_i \;\rangle   &   \\
   &  \langle \; q_i \; \neq \; c_i \;\rangle   &   \\
   &  \critone  &
\end{eqnarray*}
where $q_i$ is one of the declared parameters
and $c_i$ is one of the values in the range of $q_i$.
The values $\pil(\langle q_i = c_i \rangle )$ and
$\pil(\langle q_i \neq c_i \rangle )$
are defined by the convention at the beginning of the previous paragraph.
The value $ \pil(\critone) $ contains only the set $P(\param)$ itself.

The present definition of $\pil(\criter)$ differs slightly
from the definition of the ``pile assigned to~\criter''
in the previous design of the language~\cite{notation};
namely, we do not require that $\emptyset \in \pil(\criter) $
and $P(\criter) \in \pil(\criter) $.
We find the present definition technically more convenient.

Using these primitive criteria and the \union\ and \cross\ operations,
the test designer can write down many other useful criteria.
In particular, it is possible to specify that a particular
vector \vvx\ of parameter values $ v_i \in Q_i $ must be included
in the selected test set.
For example, to ensure that the vector in which
\begin{eqnarray*}
    \sepa \; =\; \sepb \; & = & \; \vslash  \\
    \stra \; & = & \; \vabcd                \\
    \strb \; & = & \; \vab                  \\
    \aoccurs \; & = & \; \true
\end{eqnarray*}
is in the selected set, the test designer would use the criterion
\begin{eqnarray*}
   & & \langle \; \sepa = \vslash \;\rangle  \;\;\cross\;\;
       \langle \; \sepb = \vslash \;\rangle  \;\;\cross\;\;
       \langle \; \stra = \vabcd \;\rangle  \;\;\cross\\
   & & \langle \; \strb = \vab   \;\rangle  \;\;\cross\;\;
       \langle \; \aoccurs = \true \;\rangle         \;\;   .
\end{eqnarray*}
The criteria \ea\ and \exh,
which were informally described in the previous section,
can also be constructed using \union\ and \cross.
The general definition is as follows:
Let $Q_i$ be the range of the parameter $q_i$.
If $Y \subseteq Q_i$ then define
\[
    \ea ( \; q_i \; : \; Y \;)
    \; = \; \fromto{6mm}{\bigunion}{a \in Y}{}
            \langle q_i = a \rangle \; .
\]
The criterion specifies that each value in $Y$ must be tested
(as long as there is at least one point in $D(\constr)$ with
that value of $q_i$).

For any sequence $q_{i_1} , q_{i_2} , \ldots , q_{i_m} $ of parameters,
define
\[
\exh( q_{i_1} , q_{i_2} , \ldots , q_{i_m} )
\; = \; \fromto{6mm}{\bigcross}{j=1}{m}
        \ea( \; q_{i_j} \; : \; Q_{i_j} \; )  \; .
\]
This specifies that all the combinations of values
of $q_{i_1} , q_{i_2} , \ldots , q_{i_m} $
allowed by the constraint must be tested.

The criteria \exh\qvx\ and \critone\ are at opposite ends
of the scale ordered by \wleq.
Only the set $D(\constr)$ itself is adequate for \exh\qvx.
If $D(\constr)\neq\emptyset$,
any nonempty subset of $D(\constr)$ is adequate for \critone.

\section{Worst-case complexity of two test selection problems}
    \label{sec:negative}

In this section we work with the concrete language from
Section~\ref{subsec:enum},
and we consider algorithmic aspects of the criteria specified
in the language:
Given one such criterion, how difficult is it to find an adequate test set
that is in some sense ``small''?

\subsection{Two basic problems}
    \label{sec:twobasic}

Let $ I = (\param, \constr, \criter) $ be an instance
of the test selection problem, and let $T$ be an adequate test set
for $I$.
Say that $T$ is a {\bf minimum adequate test set\/} if no set
of cardinality smaller than $|T|$ is adequate.
Say that $T$ is a {\bf minimal adequate test set\/} if no proper
subset of $T$ is adequate.

We are interested in algorithms for two problems:

\sprobl{The Minimum Adequate Set Search Problem (\MumAS)
     }{An instance $I = (\param,\constr,\criter)$.
     }{A minimum adequate test set for $I$.
     }

\sprobl{The Minimal Adequate Set Search Problem (\MalAS)
     }{An instance $I = (\param,\constr,\criter)$.
     }{A minimal adequate test set for $I$.
     }

The {\bf size of the instance\/}
$I=(\param,\constr,\criter)$, denoted \siz{I},
is the total length of the declarations in
\param\ and of the expressions \constr\ and \criter.
Often the cardinality of the set~$P(\param)$ is exponential in the
number of parameters in \param. For example, if each parameter
range~$Q_i \,$, $1 \leq i \leq N$, consists of two values then the
cardinality of $P(\param)$ is $2^N$. Thus the cardinality of a
minimum or minimal adequate test set $T$ may be exponential in
\siz{I}; in that case no algorithm that outputs $T$ can execute in
time polynomial in \siz{I}. We shall therefore measure the
execution time of such algorithms in terms of $ \siz{I} + |T| $.
Thus a {\em polynomial-time algorithm\/} for \MumAS\ or \MalAS\ is
an algorithm whose worst-case execution time is bounded by a
polynomial function of $ \siz{I} + |T| $.

We shall identify two obstacles on the path toward efficient
algorithms for \MumAS\ and \MalAS.
One obstacle, related to the boolean satisfiability problem,
applies to both \MumAS\ and \MalAS\ (section~\ref{subsec:satis});
the other, related to graph colorability, applies only to
\MumAS\ (section~\ref{sec:color}).

\subsection{Connections with boolean satisfiability}
     \label{subsec:satis}

For classifying problems as NP-complete, NP-hard, etc.,
we use the terminology of Garey and Johnson~\cite{GJ}.
\MumAS\ and \MalAS\ are {\em search\/} problems (\cite{GJ}, p.~110).
The following {\em decision\/} problem will be useful in our analysis
of the complexity of \MumAS\ and \MalAS.

\probl{The Empty Adequate Set Problem (\EA)
     }{An instance $I$.
     }{Is the empty set adequate for $I$?
     }
\\
Denote by $ \bigcup \pil(\criter) $ the union of all sets in
\pil(\criter).
The empty set is adequate for $I = (\param,\constr,\criter)$
if and only if
\( D(\constr) \cap \bigcup \pil(\criter) = \emptyset \).

It is not difficult to prove that \EA\ is in co-NP.
However, we are more interested in proving that \EA\ is NP-hard;
we now prove the NP-hardness of \EA,
by reduction from the boolean satisfiability problem.

\begin{theorem}
   \label{th:EAShard}
The problem \EA\ is NP-hard,
even if the input $I = (\param,\constr,\criter)$ is such that
\begin{enumerate}
   \item
   $\constr = \true$, or
   \item
   $\criter = \critone $.
\end{enumerate}
\end{theorem}

{\em Proof.}
By reduction from \iiiSAT\ (\cite{GJ}, p.~46).
Let $C$ be an instance of \iiiSAT.
We construct an instance $I$ such that
$C$ is satisfiable if and only if $\emptyset$ is not adequate for $I$.

Let $ C = \{ c_1 , c_2 , \ldots , c_m \} $ be a set of clauses
on a finite set $U$ of boolean variables, such that
\[
 c_j = a_{j1} \vee a_{j2} \vee a_{j3}
\]
for $1 \leq j \leq m$.
Each literal $a_{jk}$ is either a variable $u$ in $U$
or its negation $\overline{u}$.
Let \param\ be the declarations
\[
   u : \{ \; \true, \false \; \}
\]
for $u$ in $U$.

For $\constr=\true$, the empty set is adequate if and only if
\( \;\bigcup \pil(\criter) = \emptyset \).
Define
\[
    \criter = \fromto{7mm}{\bigcross}{j=1}{m}
              ( \criter_{j1} \union \criter_{j2} \union \criter_{j3} )
\]
where
\begin{eqnarray*}
\criter_{jk} & = & \left\{ \begin{array}{ll}
                              \langle u = \true \rangle
                                 & \mbox{if $ a_{jk} = u $ } \\
                              \langle u = \false \rangle
                                 & \mbox{if $ a_{jk} = \overline{u} $ }
                         \end{array}
                 \right.
\end{eqnarray*}
for $k=1,2,3$, and define
$I = (\param,\true,\criter)$.
Then $C$ is satisfiable if and only if
$ \;\bigcup \pil(\criter) \neq \emptyset $.

For $\criter=\critone$,
the empty set is adequate if and only if \( D(\constr) = \emptyset \).
Define
\[
     \constr = \fromto{6mm}{\bigwedge}{j=1}{m}
     (\constr_{j1} \vee \constr_{j2} \vee \constr_{j3} )
\]
where
\begin{eqnarray*}
\constr_{jk} & = & \left\{ \begin{array}{ll}
                              u = \true
                                 & \mbox{if $ a_{jk} = u $ } \\
                              u = \false
                                 & \mbox{if $ a_{jk} = \overline{u} $ }
                         \end{array}
                 \right.
\end{eqnarray*}
and define $I = (\param,\constr,\critone)$.
Then $C$ is satisfiable if and only if
\( D(\constr) \neq \emptyset \).
\qed

The following lemma shows that any lower bound for the execution
time complexity of \EA\ implies a lower bound for \MumAS\ and \MalAS.

\begin{lemma}
    \label{prop:MAStoEAS}
Let $w$ be an integer function of an integer variable
such that the value $w(i)$
for any integer $i$ can be computed in $O(w(i))$ steps.
If there exists an algorithm for \MumAS\ or \MalAS\ that
for every input $I$ produces an output $T$
in at most $w(\siz{I}+|T|)$ steps,
then there exists an algorithm that solves \EA\ for every input $I$
in $O(w(\siz{I}))$ steps.
\end{lemma}

{\em Proof.}
To solve \EA\ on input $I$,
compute $w(\siz{I})$ and execute the algorithm for
\MumAS\ (or \MalAS) on input~$I$ for at most $w(\siz{I})$ steps.
The answer to the question in \EA\ is ``yes''
if the algorithm terminates with output $T=\emptyset$.
The answer is ``no'' if the algorithm terminates with output
$T\neq\emptyset$ or does not terminate in $w(\siz{I})$ steps.
\qed

\begin{theorem}
If P $\neq$ NP then neither \MumAS\ nor \MalAS\ is solvable
by a polynomial-time algorithm,
even in cases~1 and~2 in Theorem~\ref{th:EAShard}.
\end{theorem}

{\em Proof.}
Apply Theorem~\ref{th:EAShard} and Lemma~\ref{prop:MAStoEAS}.
\qed

\subsection{Connections with graph colorability}
     \label{sec:color}

We have identified one reason why \MumAS\ and \MalAS\ are difficult:
\constr\ and \criter\ may encode arbitrary boolean expressions,
and thus any algorithm for \MumAS\ or \MalAS\
can be used to construct an algorithm for \iiiSAT.
It is therefore natural to ask whether \MumAS\ and \MalAS\ become
easier when \constr\ and \criter\ belong to a smaller class
of expressions.

We start with a simple such class, the criteria in \uce\ form.
The {\bf \uce\ form} of a test selection criterion is
\begin{equation}
   \fromto{6mm}{\bigunion}{j=1}{m}  \;
   \fromto{6mm}{\bigcross}{k=1}{n_j} \;
   \;\criter_{jk}
                         \label{eq:uceform}
\end{equation}
where $\criter_{jk} $ are primitive criteria
of the form $ \langle q_i = c_i \rangle $.
Define an instance $I = (\param,\constr,\criter)$
to be {\bf simple\/} if $\constr = \true $ and \criter\ is in \uce\ form.
In the next section we shall see that the problem \MalAS\
for simple instances is solvable by a polynomial-time algorithm.
In contrast, \MumAS\ for simple instances is NP-hard,
as will be established in Theorem~\ref{th:MumASsimple}.
The following decision problem will be used in the proof.

\probl{The Minimum Adequate Set Problem for Simple Instances (\MASI)
     }{A simple instance $I$ and an integer $K$.
     }{Is there a set $T$ adequate for $I$ such that $ |T| \leq K $ ?
     }
\\
We are going to show that \MASI\ is equivalent to \GRAPHCOL\
(\cite{GJ}, p.~191).

Let \pil\ be a set of sets.
The {\bf intersection graph\/} of \pil\ is the graph $ G = (\pil,E) $
in which the set of vertices is \pil\ and the set of edges is
\[
E = \{ \; \{ X, Y \} \;\; |
\;\; X, Y \in \pil, \; X \neq Y \et X \cap Y \neq \emptyset \; \} \; .
\]

When $ G = ( V , E ) $ is a graph, the {\bf complement\/} of $G$
is the graph $ \overline{G} = ( V , \overline{E} ) $, where
\[
   \overline{E} = \{ \; \{ x,y\} \;\; | \;\; x,y \in V,
       \; x \neq y \;\;\mbox{\rm and}\;\; \{ x,y\} \not\in E \; \} \;\; .
\]

The proof of the following simple lemma is left to the reader.
Note that the lemma would not be true if we admitted primitive
criteria of the form $ \langle q_i \neq c_i \rangle $.

\begin{lemma}
     \label{prop:pairwise}
Let $I = (\param,\true,\criter)$ be a simple instance, and let
$ \pil_0 \subseteq \pil(\criter) $.
If $ X \cap Y \neq \emptyset $ for all $ X, Y \in \pil_0 $
then $ \bigcap \{ X | X \in \pil_0 \} \neq \emptyset $.
\qed
\end{lemma}

By the lemma, a set $ \pil_0 \subseteq \pil(\criter) $
forms a clique in the intersection graph of \pil(\criter)
if and only if $ \bigcap \{ X | X \in \pil_0 \} \neq \emptyset $.

\begin{proposition}
    \label{prop:color}
Let $I = (\param,\true,\criter)$ be a simple instance,
let $\overline{G}$ be the complement of the intersection graph of
$\pil(\criter) \setminus \{ \emptyset \}$,
and let $K$ be an integer.
The graph $\overline{G}$ is $K$-colorable if and only if
there exists a set $T$ adequate for $I$ such that $ |T| \leq K $.
\end{proposition}

{\em Proof.}
Assume
$\overline{G} = (\pil(\criter)\setminus\{\emptyset\}, \overline{E})$
is $K$-colorable.
This means that there exists a mapping
$f : \pil(\criter) \setminus \{ \emptyset \} \rightarrow
\{1,2,\ldots,K\}$ such that
$ f(X) \neq f(Y) $ when $ \{ X, Y \} \in \overline{E} $.
Define
\[
   \pil_j = \{ \; X \in \pil(\criter) \setminus \{ \emptyset \} \; |
               \; f(X) = j \; \}
\]
for $ j = 1,2,\ldots,K$.
From the definition of the intersection graph we get that
if $X,Y \in \pil_j$ then $X \cap Y \neq \emptyset $;
by Lemma~\ref{prop:pairwise}
we have $ \bigcap \{ X | X \in \pil_j \} \neq \emptyset $.
Form a set $T$ by choosing one point in each
$ \bigcap \{ X | X \in \pil_j \} $, $j=1,2,\ldots,K$.
Thus $|T| \leq K$ and $T$ intersects each nonempty $X\in \pil(\criter)$,
which means that $T$ is adequate for $I$.

Conversely, assume that
there exists a set $T$ adequate for $I$ such that $ |T| \leq K $.
Write $ T = \{ d_1 , d_2 , \ldots , d_K \} $ and define
a $K$-coloring
$f : \pil(\criter) \setminus \{ \emptyset \} \rightarrow
\{1,2,\ldots,K\}$
of $\overline{G}$ by
\[
     f(X) = \min \{ \; j \; | \; d_j \in X \; \} \; .
\]
Since $T$ is adequate, $f$ is defined for each
$ X \in \pil(\criter) \setminus \{ \emptyset \} $.
If $ f(X) = f(Y) = j $ then  $ d_j \in X \cap Y $,
hence $ X \cap Y \neq \emptyset $,
hence $ \{ X , Y \} $ is not an edge in $\overline{G}$.
Thus $f$ is a coloring of~$\overline{G}$.
\qed

\begin{proposition}
     \label{prop:findinst}
For each graph $G=(V,E)$ there exists a simple instance
$I = (\param,\true,\criter)$ such that the intersection graph of
$\pil(\criter)$ is (isomorphic to) $G$.
The declarations \param\ consist of one boolean parameter
for each vertex in $V$.
\end{proposition}

{\em Proof.}
Let $V$ consist of $N$ vertices, $ V = \{ x_1 , x_2 , \ldots , x_N \} $.
Let \param\ be the declarations
\[
    q_i : \{ \true , \false \}
\]
for $ i = 1,2, \ldots , N $.
Define
\[
    H(i) = \{ \; j \in \{1,2,\ldots,N\} \;\; | \;\; j \neq i
         \et \{ x_i , x_j \} \not\in E \; \} \;\; ,
\]
\[
    \criter \; = \;
    \fromto{6mm}{\bigunion}{i=1}{N} \;
      ( \;\langle q_i = \true \rangle \;\cross\;
    \fromto{11mm}{\bigcross}{j \in H(i)}{}
          \langle q_j = \false \rangle\;
      )  \;\; .
\]
Then \pil(\criter) consists of the sets
\[
     X_i = \{ \; \vvx \in P(\param) \;\; | \;\; v_i = \true \et
              \forall j \in H(i) \; : \; v_j = \false \; \}
\]
for $ i=1,2,\ldots,N$.
We have $ X_i \neq X_j $ for $ i \neq j $,
the sets $ X_i $ are nonempty,
and the mapping $ x_i \mapsto X_i \;$ is an isomorphism between
$G$ and the intersection graph of \pil(\criter).
\qed

By Propositions~\ref{prop:color} and~\ref{prop:findinst},
\GRAPHCOL\ and \MASI\ are polynomially equivalent.
From known results for \GRAPHCOL\ (\cite{GJ}, p.~191)
we obtain the following result for \MASI.

\begin{theorem}
    \label{th:MASINPcompl}
The problem \MASI\ is NP-complete, even for $K=3$.
\qed
\end{theorem}

It remains to transform \MASI\ into \MumAS.
The only potential complication is that ``polynomial''
means ``polynomial in the size of input'' for \MASI\
and ``polynomial in the size of input {\em and output\/}''
for \MumAS.
However, if the input instance $I$ is simple
and the output set $T$ is minimum then $|T|$ is bounded
by \siz{I}.
Indeed, for criterion~(\ref{eq:uceform})
there exists an adequate test set of cardinality at most $m$,
which means that the cardinality of the minimum set $T$
is also bounded by $m$.
Thus Theorem~\ref{th:MASINPcompl} yields the following
result for \MumAS.

\begin{theorem}
   \label{th:MumASsimple}
If P $\neq$ NP then \MumAS\
is not solvable by a polynomial-time algorithm,
even for simple instances.
\qed
\end{theorem}

In view of Theorem~\ref{th:MumASsimple},
we are not likely to find a polynomial-time algorithm for \MumAS.
It is still possible that there is an algorithm for \MumAS\
that is efficient in some other sense, but we have not been able
to find any such algorithm.
However, in the next section we present a practical
algorithm for \MalAS.

The transformation in
Propositions~\ref{prop:color} and~\ref{prop:findinst}
yields more than results for \MASI\ and \MumAS.
For example, if we had an algorithm that for every simple instance
would find an adequate test set whose cardinality is within the factor
$(1+\varepsilon)$ of the minimum,
then we would also have an algorithm to color any graph with
the number of colors within the factor $(1+\varepsilon)$ of the minimum.
No such polynomial-time algorithm is presently known for any fixed
constant $\varepsilon$.

By virtue of Proposition~\ref{prop:color},
any algorithm for graph coloring can be transformed into an algorithm
for constructing adequate test sets for simple instances;
when the graph coloring uses the minimum number of colors,
the adequate test set is minimum.
Many heuristic algorithms for graph coloring have been studied;
see e.g.~\cite{philips,toft} and the references therein.
However, we are interested in the problems \MumAS\ and \MalAS,
rather than \MASI;
the restriction to simple instances is severe.
We have already noted that Lemma~\ref{prop:pairwise} does not hold
if primitive criteria $ \langle q_i \neq c_i \rangle $ are allowed.
Moreover, if \constr\ is a general constraint then
Lemma~\ref{prop:pairwise} may fail for the sets
in \pil(\criter) restricted to the domain $D(\constr)$.

\section{Algorithms for finding minimal adequate test sets}
         \label{sec:positive}

\subsection{An algorithm for normalized instances}

In this section we concentrate on the problem \MalAS\ defined
in Section~\ref{sec:twobasic}.
We start with an efficient algorithm for the input instances
$ I=(\param,\constr,\criter) $ in which \constr\ and
\criter\ belong to a certain restricted class of expressions.
Afterwards we show how to use the algorithm for general instances.

The test selection criterion
\begin{equation}
   \fromto{6mm}{\bigunion}{j=1}{r}   \;
   \fromto{6mm}{\bigcross}{k=1}{s_j} \;\;   \criter_{jk} \;\; ,
                         \label{eq:ucform}
\end{equation}
where $\criter_{jk} $ are primitive criteria,
is said to be in the \union\cross\ form.
The constraint
\begin{equation}
   \fromto{6mm}{\bigvee}{j=1}{m}   \;
   \fromto{6mm}{\bigwedge}{k=1}{n_j} \;\;   \constr_{jk} \;\; ,
                         \label{eq:vwform}
\end{equation}
where $\constr_{jk} $ are primitive constraints,
is said to be in the $\vee\wedge$ form.
(This is also called the disjunctive normal form.)

An instance $I=(\param,\constr,\criter)$
is {\bf normalized\/} if \constr\ is in the $\vee\wedge$~form and
\criter\ in the \union\cross~form.

Let \param\ be a fixed set of parameter declarations
$ \; q_i : Q_i $, $\; i=1,2,\ldots,N$.
We say that a set $ X \subseteq P(\param) $ is a {\bf subcube\/}
if it is in the form
\(
    \prod_{i=1}^N R_i
\)
where $ R_i \subseteq Q_i $.
For our concrete language of Section~\ref{subsec:enum},
every $ X \in \pil(\criter) $ is a subcube.
When the criterion \criter\ is in the \union\cross\ form,
it is easy to compute the set \pil(\criter):
The subcubes in \pil(\criter) correspond to the terms
$\bigcross_k \criter_{jk}$ in~(\ref{eq:ucform}).
Similarly, every term
$\bigwedge_k \constr_{jk}$ in~(\ref{eq:vwform})
defines the subcube
$D(\bigwedge_k \constr_{jk})$, and
\(
   D ( \bigvee_j \bigwedge_k \; \constr_{jk}  )
  =    \bigcup_j \;
   D ( \bigwedge_k \; \constr_{jk} )
\).

The algorithm in Figure~\ref{fig:algMal} constructs a minimal
set adequate for a given normalized instance.
The input for the algorithm consists of two sets of subcubes:
the set $S=\pil(\criter)$, and the set
\[
  {\textstyle
  C = \{ \; D(\bigwedge_k \; \constr_{jk}) \;
       | \; j=1,2,\ldots,m \; \}
  }
\]
for the constraint~(\ref{eq:vwform}).
When the algorithm terminates,
the set variable $T$ contains a minimal adequate set.

\begin{figure}[tb]
\fuline
\begin{tabbing}
\hspace{6mm}\=\hspace{6mm}\=\hspace{6mm}\=\hspace{6mm}\=\hspace{6mm}\=
\hspace{6mm}\=\hspace{6mm}\=\hspace{6mm}\=\hspace{6mm}\=\hspace{6mm}\=
                                                             \kill
{\bf inputs}
\\[-2mm] \>$S$ \>: set of subcube
\\[-2mm] \>$C$ \>: set of subcube
\\[-2mm]
                                                              \\[-2mm]
{\bf variables}
\\[-2mm] \>$T$ \>\>\>\ \ : set of point
\\[-2mm] \>$contains(t)$ \>\>\>\ \ : set of subcube, \ \ for $t
\in T$ \\[-2mm] \>$count(X)$ \>\>\>\ \ : integer, \ \ for $X \in
S$           \\[-2mm]
                                                              \\[-2mm]
{\bf initially}
\\[-2mm] \>$T = \emptyset$
\\[-2mm] \>$count(X) = 0$, \ \ for $X \in S$
\\[-2mm]
                                                              \\[-2mm]
{\bf program}
\\[-2mm] \>{\bf forall} \ \ $X \in S$ \ \ {\bf do}
\\[-2mm] \>\>{\bf if} \ \ $count(X) = 0$ \ \ {\bf then}
\\[-2mm] \>\>\>$t$ := Find\_point( $X$, $C$ )
\\[-2mm] \>\>\>{\bf if} \ \ $t \neq NIL$ \ \ {\bf then}
\\[-2mm] \>\>\>\>$T := T \cup \{ t \}$
\\[-2mm] \>\>\>\>{\bf forall} \ \ $Y \in S$ \ \ {\bf do}
\\[-2mm] \>\>\>\>\>{\bf if} \ \ $ t \in Y$ \ \ {\bf then}
\\[-2mm] \>\>\>\>\>\>$contains(t) := contains(t) \cup \{ Y \}$
\\[-2mm] \>\>\>\>\>\>$count(Y) := count(Y) + 1$
\\[-2mm]
                                                              \\[-2mm]
\>{\bf forall} \ \ $t \in T$ \ \ {\bf do}
\\[-2mm] \>\>{\bf if} \ \ $\max(\; count(Y),
  \; Y \!\in\! contains(t)\; ) \geq 2$ \ \ {\bf then}         \\[-2mm]
\>\>\>{\bf forall} \ \ $Y \in contains(t)$ \ \ {\bf do}
\\[-2mm] \>\>\>\>$count(Y) := count(Y) - 1$
\\[-2mm] \>\>\>$T := T \setminus \{ t \}$
\\[-2mm]
\end{tabbing}
\fuline
\caption{Algorithm for \MalAS}
\label{fig:algMal}
\end{figure}

In the program for the algorithm, {\bf forall} denotes iteration
over all elements of a set in some arbitrary order.
The values of the data type ``point'' are the elements of $P(\param)$.
The function call Find\_point($X,C$) finds a point
in the set $ X \cap \bigcup C $; if the set is empty,
the function returns NIL.

For each $t\in T$, the variable $contains(t)$ stores a set of subcubes;
a subcube $X\in S$ belongs to $contains(t)$ if and only if $t \in X$.
For each $X \in S$,
the variable $count(X)$ stores the cardinality of $X \cap T$.

The algorithm works in two phases: The first phase finds an
adequate test set, and the second phase trims the set to make it
minimal.

When sets are represented as arrays or linked lists,
adding one element takes constant time,
and iterating through a {\bf forall\/} loop
adds only constant time per iteration.
The deletion operation on the last line of the program
is implemented by marking the element as deleted;
that also takes only constant time.

When points and subcubes are represented as sorted lists of
primitive constraints, the function Find\_point($X,C$) and the
test ``{\bf if } $t\in Y$'' are implemented by a single pass
through the lists representing the two arguments. Adding it all
up, we get the bound $O(|I|^2)$ for the total execution time of
the algorithm on any input instance $I$. We summarize our analysis
in a theorem.

\begin{theorem}
There is an algorithm to solve the problem \MalAS\ for any
normalized instance $I$ in time $O(|I|^2)$. \qed
\end{theorem}

\subsection{The cost of normalization}

Requiring input instances to be normalized would be
inconvenient to the users.
For example:
\begin{itemize}
\item
The constraint is often naturally specified in the conjunctive,
rather than disjunctive, normal form.
\item
The user should be able to
take any two criteria and combine them by means of \cross.
The resulting criterion is not in the \union\cross\ form.
\end{itemize}
Therefore our design allows users to specify the instance
in the general form defined in Section~\ref{sec:nota}.
The instance is automatically converted into an equivalent normalized form
before the algorithm in Figure~\ref{fig:algMal} is applied.

The normalization is easy to implement.
The well-known procedure
transforms boolean expressions into $\vee\wedge$ form by
repeatedly replacing a conjuction of disjunctions by an equivalent
disjunction of conjunctions.
By virtue of the distributive law for
\union\ and \cross, the same procedure works for the test
selection criteria built using \union\ and \cross.

However, the user should understand that the normalization
may in some cases be expensive, in terms of execution time.
In the worst case, the execution time is exponential in the size
of the original expression.
We shall now discuss the implications of the normalization cost,
separately for the constraint expression \constr\ and for the criterion
expression~\criter.

For \constr, the exponential increase of the execution time of the
normalization procedure is more common and more serious than for
\criter. A large instance $ I=(\param,\constr,\criter) $ of the
test selection problem is typically obtained by putting together
several instances $ I_j =(\param_j ,\constr_j ,\criter_j ) $ with
disjoint sets of parameters. It is then natural to take $\constr =
\bigwedge_j \constr_j $.
If $m$ independent constraints are put
together to form
\[
   \constr = \bigwedge_{j=1}^m \; ( \constr_{j1} \vee \constr_{j2} )
\]
then the equivalent $\vee\wedge$ form of \constr\ has $2^m$ terms.
Thus in this case the total execution time is at least
proportional to $2^m$, even if the test set produced at the end is
very small.

The cost of normalizing the criterion \criter\ is less critical;
in most cases large criteria lead to large test sets.
However, ``in most cases'' does not mean ``always'', as the following
example shows:
\newline
{\bf Example.}
Let \param\ consist of $N$ declarations $ q_i : \{ 0 , 1 \} $,
$ 1 \leq i \leq N $.
Consider the criterion
\begin{equation}
    \fromto{6mm}{\bigcross}{j=1}{N}
    \mbox{}
    \fromto{6mm}{\bigunion}{i=1}{N}
    \; \langle q_i = 0 \rangle \;\; .
           \label{eq:before}
\end{equation}
The equivalent \union\cross\ form is
\begin{equation}
    \fromto{6mm}{\bigunion}{A}{}
    \mbox{}
    \fromto{6mm}{\bigcross}{i\in A}{}
    \; \langle q_i = 0 \rangle \;\; ,
           \label{eq:after}
\end{equation}
where $A$ runs through all nonempty subsets of $\{ 1,2,\ldots,N
\}$. The only minimal adequate test set is $ T = \{
(0,0,\ldots,0)\} $, of cardinality 1. In transforming
(\ref{eq:before}) to (\ref{eq:after}) the algorithm generates all
the $2^N -1$ expressions
\[
    \fromto{6mm}{\bigcross}{i\in A}{}
    \; \langle q_i = 0 \rangle \;\; ,
\]
where $ \emptyset \neq A \subseteq \{ 1,2,\ldots,N \}$.
\qed
\newline
Nevertheless, we conjecture that, in most practical situations,
if the test designer specifies a selection criterion
whose equivalent \union\cross\ form is very large,
then every adequate test set will also be very large.
In such cases long execution time
(at least proportional to the size of the produced test set)
cannot be avoided.

In the next section we shall show that
for the instance $I$ built by combining independent instances $I_j$,
we can solve the problem \MalAS\ separately for each $I_j$ and then
put the solutions together to produce a test set adequate for $I$.
We shall also describe an algorithm for decomposing
instances into independent components.
We expect that for most instances of the test selection problem
arising in practice the decomposition method will avoid
the exponential cost of normalization.

\subsection{Decomposition of instances}
      \label{sec:decomposition}

When the test designer constructs a large instance of
the test selection problem, it is likely that the instance
is built from subproblems that are in some sense independent.
Now we show how such structure can be exploited
to construct minimal adequate test sets.

{\bf Definition.}
Two instances
$ I_j = ( \param_j, \constr_j, \criter_j ) $, $ j=1,2$,
are {\bf independent} if no parameter occurs in both
$\param_1$ and $\param_2$.

{\bf Definition.}
Let $ I_j = ( \param_j, \constr_j, \criter_j ) $, $ j=1,2$,
be two independent instances.
Let $ \param = \param_1 \cup \param_2$.
Define two instances
\begin{eqnarray*}
    I_1 [ \crossand ] I_2 & = &
    ( \param, \constr_1 \wedge \constr_2,
      \criter_1 \cross \criter_2 )              \\
    I_1 [ \unionand ] I_2 & = &
    ( \param, \constr_1 \wedge \constr_2,
      \criter_1 \union \criter_2 )
\end{eqnarray*}
When $\alpha$ is \crossand\ or \unionand, we say that $ I_1 $ and
$I_2$ form an {\em independent $\alpha$-decomposition\/} (or
simply a {\em decomposition\/}) of $ I_1 [ \alpha ] I_2 $.

We now construct adequate test sets for $ I_1 [ \crossand ] I_2 $
and $ I_1 [ \unionand ] I_2 $ from adequate test sets for $I_1$
and $I_2$. For two nonempty sets $T_1$ and $T_2$ such that $|T_1 |
= m$, $|T_2 | = n $, define the set $ T_1 \eqmerge T_2 \subseteq
T_1 \times T_2$ as follows: Let $T_1 = \{ r_1 , r_2 , \ldots , r_m
\}$, $T_2 = \{ s_1 , s_2 , \ldots , s_n \}$, and \[
       T_1 \eqmerge T_2
            =  \left\{ \begin{array}{ll}
                        \{ \;(r_1 , s_1 ), (r_2 , s_2 ), \ldots ,
                           (r_m , s_m )\;\} & \mbox{if $m = n$}  \\
                        \{ \;(r_1 , s_1 ), (r_2 , s_2 ), \ldots ,
                           (r_n , s_n ), (r_{n+1} , s_n), \ldots ,
                           (r_m , s_n) \;\}    & \mbox{if $m > n$}  \\
                        \{ \;(r_1 , s_1 ), (r_2 , s_2 ), \ldots ,
                           (r_m , s_m ), (r_m, s_{m+1}), \ldots ,
                           (r_m, s_n ) \;\}    & \mbox{if $m < n$}
                      \end{array}
              \right.
\]
Thus the definition of $ T_1 \eqmerge T_2 $ depends on the order in
which we number the elements of $T_1$ and $T_2$;
we assume that one such order is chosen arbitrarily.

{\bf Definition.}
Let $ I_j = ( \param_j, \constr_j, \criter_j ) $, $ j=1,2$,
be two independent instances such that $D(\constr_j) \neq \emptyset$,
and let $T_j \subseteq D(\constr_j)$ for $j=1,2$.
Define
\begin{eqnarray*}
     T_1 [ \crossand ] T_2  & = & T_1 \times T_2  \\
     T_1 [ \unionand ] T_2  & = &
              \left\{ \begin{array}{ll}
                         \emptyset
                           & \mbox{if $T_1 = \emptyset = T_2$} \\
                         T_1 \times \{ s_2 \}
                           & \mbox{if $T_1 \neq \emptyset = T_2$} \\
                         \{ s_1 \} \times T_2

                           & \mbox{if $T_1 = \emptyset \neq T_2$} \\
                         T_1 \eqmerge T_2
                           & \mbox{if $T_1 \neq \emptyset \neq T_2$}
                         \end{array}
                 \right.
\end{eqnarray*}
where $ s_j \in D(\constr_j) $, $j=1,2$,
are some arbitrarily chosen elements.

\begin{theorem}
    \label{th:composition}
Let $ I_j = ( \param_j, \constr_j, \criter_j ) $, $ j=1,2$, be two
independent instances such that $D(\constr_j ) \neq \emptyset$,
and let $\alpha$ be \crossand\ or \unionand. If $ T_j$ is an
adequate test set for $I_j$, $j=1,2$, then $ T_1 [ \alpha ] T_2 $
is an adequate test set for $ I_1 [ \alpha ] I_2 $. If $ T_j $ is
a minimal adequate test set for $I_j$, $j=1,2$, then $ T_1 [
\alpha ] T_2 $ is a minimal adequate test set for $ I_1 [ \alpha ]
I_2 $.
\end{theorem}

{\em Proof.}
Let $ \param\ = \param_1 \cup \param_2$,
$ P_1 = P(\param_1 ) $,
$ P_2 = P(\param_2 ) $, $ P = P(\param) $,
and $ \constr = \constr_1 \wedge \constr_2 $.
Thus
\begin{eqnarray*}
     P & = & P_1 \times P_2                                  \\
     D( \param, \constr ) & = & D(\param_1 , \constr_1 )
                             \times D(\param_2 , \constr_2 )     \\
     \pil( \param , \criter_1 ) & = &
             \{ \; X_1 \times P_2 \;
             | \; X_1 \in \pil( \param_1 , \criter_1 ) \; \}    \\
     \pil( \param , \criter_2 ) & = &
             \{ \; P_1 \times X_2 \;
             | \; X_2 \in \pil( \param_2 , \criter_2 ) \; \}    \\
     \pil(\param, \criter_1 \union \criter_2 ) & = &
       \pil(\param, \criter_1 ) \;\cup\; \pil(\param, \criter_2 )  \\
     \pil(\param, \criter_1 \cross \criter_2 ) & = &
             \{ \; X_1 \times X_2 \;
             | \; X_1 \in \pil(\param_1 , \criter_1 ) \; , \;
               X_2 \in \pil(\param_2 , \criter_2 ) \; \}
\end{eqnarray*}

Let $T_j$ be an adequate test set for $I_j$, $j=1,2$; that is, $
T_j \cap X \neq \emptyset $ whenever $ X \in \pil(\param_j ,
\criter_j ) $ and $ X \cap D(\param_j , \constr_j ) \neq \emptyset
$. Let $ T = T_1 [ \alpha ] T_2 $.

For $ \alpha = \crossand $,
if $ X_1 \times X_2 \in \pil(\param, \criter_1 \cross \criter_2 )$
and $ ( X_1 \times X_2 ) \cap D(\param, \constr ) \neq \emptyset $
then $ X_j \in \pil(\param_j , \criter_j )$,
$ X_j \cap D(\param_j , \constr_j ) \neq \emptyset $.
Therefore $ X_j \cap T_j \neq \emptyset $ and
$( X_1 \times X_2 ) \cap T \neq \emptyset $.

For $ \alpha = \unionand $,
if $ X_1 \times P_2 \in \pil(\param, \criter_1 )$
and $ ( X_1 \times P_2 ) \cap D(\param, \constr ) \neq \emptyset $
then $X_1 \in \pil(\param_1 , \criter_1 )$,
$ X_1 \cap D(\param_1 , \constr_1 ) \neq \emptyset $.
Therefore $ X_1 \cap T_1 \neq \emptyset $ and
$( X_1 \times P_2 ) \cap T \neq \emptyset $.
The argument for $ P_1 \times X_2 \in \pil( \param, \criter_2 ) $
is symmetrical.

Now let $T_j$ be a minimal adequate test set for $I_j$, $j=1,2$,
and let $ T = T_1 [ \alpha ] T_2 $.

Let $ \alpha = \crossand $.
To prove that $T$ is minimal, take any $ ( t_1 , t_2 ) \in T$.
Since $T_j$ is minimal, there is
$ X_j \in \pil(\param_j , \constr_j )$
such that $ X_j \cap D(\param_j , \constr_j ) \neq \emptyset $ and
$ X_j \cap \left( T_j \setminus \{ t_j \} \right) = \emptyset $.
For $ X = X_1 \times X_2 $ we have
$ X \cap D(\param , \constr ) \neq \emptyset $ and
$ X \cap \left( T \setminus \{ ( t_1 , t_2 ) \} \right) = \emptyset $.
Therefore $ T \setminus \{ ( t_1 , t_2 ) \} $ is not adequate.
Thus $T$ is minimal.

Let $ \alpha = \unionand $.
If $ T_1 = T_2 = \emptyset $ then $T=\emptyset$, hence $T$ is minimal.
Now assume, without loss of generality, that
$ | T_1 | \geq | T_2 | $ and $ T_1 \neq \emptyset $.
Then for every $t_1 \in T_1$ there exists exactly one
$t_2 \in D(\constr_2 )$ such that $ ( t_1 , t_2 ) \in T $.
To prove that $T$ is minimal, take any $ ( t_1 , t_2 ) \in T$.
Since $T_1$ is minimal, there is
$ X_1 \in \pil(\param_1 , \constr_1 )$
such that $ X_1 \cap D(\param_1 , \constr_1 ) \neq \emptyset $ and
$ X_1 \cap \left( T_1 \setminus \{ t_1 \} \right) = \emptyset $.
For $ X = X_1 \times P_2 $ we have
$ X \cap D(\param , \constr ) \neq \emptyset $ and
$ X \cap \left( T \setminus \{ ( t_1 , t_2 ) \} \right) = \emptyset $.
Therefore $ T \setminus \{ ( t_1 , t_2 ) \} $ is not adequate.
Thus $T$ is minimal.
\qed

One can prove that if $T_1$ and $T_2$ are {\em minimum\/} adequate
then $ T_1 [ \unionand ] T_2 $ is also minimum. However, the same
is not true for $ T_1 [ \crossand ] T_2 $, as the following
example shows:

{\bf Example.}
Define two instances $ I_j = ( \param_j , \constr_j , \criter_j )$,
$j=1,2$:
The declaration $\param_j$ is
\[
      x_j  \; : \; \{ 1,2,3 \} \;\; ,
\]
there is no constraint (i.e. $\constr_j = \true$),
and the criterion $\criter_j$ is
\[
      \langle x_j \neq 1 \rangle \;\union\;
      \langle x_j \neq 2 \rangle \;\union\;
      \langle x_j \neq 3 \rangle \;\; .
\]
If $T_1$ and $T_2$ are minimum adequate sets for $I_1$ and $I_2$
then $|T_1 | = |T_2 | = 2$, hence $ | T_1 \times T_2 | = 4 $.
However, the three-element set $\{ (1,1), (2,2), (3,3) \}$ is
adequate for $ I_1 [ \crossand ] I_2 $. \qed

To utilize Theorem~\ref{th:composition} in constructing minimal
test sets, we simply add the operations $ [ \crossand ] $ and $ [
\unionand ] $ on instances to the language. The test designer may
then specify a large instance as a combination of smaller
components, using $ [ \crossand ] $ and $ [ \unionand ] $. In
fact, if the language has appropriate scoping rules for the names
of parameters then we need not require that the parameter names in
the component instances be different.

Now we describe a simple algorithm for discovering a decomposition
into independent instances, when the decomposition is not explicitly
specified by the test designer.
The algorithm works on the instances $I = (\param,\constr,\criter)$
in which \constr\ has the form $ \bigwedge_k \constr_{k} $.
The algorithm groups some terms $\constr_k$ and some subexpressions
of \criter\ together, but does not attempt to use distributive laws
to transform the expressions \constr\ and \criter.

Consider an instance $I = (\param,\constr,\criter)$
in which \param\ consists of declarations $q_i : Q_i$, $i=1,2,\ldots,N$.
{\bf Subexpressions\/} (often called {\bf well-formed subexpressions\/})
of \criter\ correspond to subtrees of the parse tree of \criter.
For $i=1,2,\ldots,N$, let $\criter(i)$ be the smallest subexpression
of \criter\ that contains all occurrences of $q_i$ in \criter;
in the parse tree of \criter,
$\criter(i)$ corresponds to the smallest subtree containing
all the leaves labeled
$ \langle q_i = c \rangle $ and
$ \langle q_i \neq c \rangle $.

Define two binary relations $W_{\constr}$ and $W_{\criter}$
on the set $\{ 1,2,\ldots,N\}$:
\begin{itemize}
   \item
   $ i \;\;W_{\constr}\;\; j \;\;$
   if $i$ and $j$ occur in $\constr_k$, for some $k$;
   \item
   $ i \;\;W_{\criter}\;\; j \;\;$
   if $j$ occurs in $\criter(i)$.
\end{itemize}
Let $W$ be the finest equivalence relation on $\{ 1,2,\ldots,N\}$
such that $ W \supseteq W_{\constr} \cup W_{\criter} $.
Computing $W$ is a straightforward application of
the transitive-closure algorithm (\cite{AHU}, p.~199).
Now each equivalence class $B$ of $W$ determines a subset $\param_B$
of the declarations \param; the subsets $\param_B$ are pairwise disjoint.
By the construction of $W$ we have
\[
     \constr = \fromto{5mm}{\bigwedge}{B}{} \;
               \fromto{5mm}{\bigwedge}{k \in B}{} \;
               \constr_k
\]
where $\fromto{5mm}{\bigwedge}{B}{}$ is the conjunction over the
equivalence classes $B$ of $W$. Each equivalence class $B$
determines a subexpression $\criter_B$. The expression \criter\ is
formed from $\criter_B$ by means of \union\ and \cross. Thus we
have decomposed $I$ into independent instances $I_B$, from which
$I$ is formed by means of $[\unionand]$ and $[\crossand]$.

It is of course possible that $i \, W \, j$ for all $i,j \in \{
1,2,\ldots,N\}$. In that case this simple approach to
decomposition does not help. However, in those cases where $I$ has
been formed by combining several independent instances using
$[\unionand]$ and $[\crossand]$, the algorithm will lead back at
least to the original independent instances, and it may even
discover a decomposition into smaller instances.

\subsection{Generalized decomposition}

In analogy to the operations $ [ \crossand ] $ and $ [ \unionand ]
$, we can also define
\begin{eqnarray*}
    I_1 [ \crossor ] I_2 & = &
    ( \param_1 \cup \param_2 , \constr_1 \vee \constr_2,
      \criter_1 \cross \criter_2 )        \\
    I_1 [ \unionor ] I_2 & = &
    ( \param_1 \cup \param_2 , \constr_1 \vee \constr_2,
      \criter_1 \union \criter_2 )
\end{eqnarray*}
whenever $ I_j = ( \param_j, \constr_j, \criter_j ) $, $ j=1,2$,
are two independent instances.

However, to construct an adequate set for $ I_1 [ \crossor ] I_2 $
or $ I_1 [ \unionor ] I_2 $, we need more than adequate sets for
$I_1$ and $I_2$. A set $T \subseteq P(\param)$ is an {\bf extended
test set\/} for $ I = ( \param, \constr, \criter ) $ if $T$ is
adequate for $ ( \param, \true, \criter ) $ and $ T \cap D $ is
adequate for $I$. The following lemma and its corollary tie
together extended and adequate sets. We leave the easy proof to
the reader.

\begin{lemma}
If $T$ and $T'$ are two extended test sets for $I$ then so is
$ ( T \cap D ) \cup ( T' \setminus D ) $.
\end{lemma}

\begin{corollary}
If T is a minimal extended test set for $I$ then $T\cap D$
is a minimal adequate set for~$I$.
\end{corollary}

The advantage of working with extended test sets is that extended
test sets for $ I_1 [ \crossor ] I_2 $ and $ I_1 [ \unionor ] I_2
$ can be constructed from extended test sets for $I_1$ and $I_2$,
and the construction preserves the property of being minimal. For
$[\crossor]$ we simply define $ T_1 [ \crossor ] T_2   =  T_1
\times T_2 $. The definition of $T_1 [\unionor] T_2$ resembles
that of $T_1 [\crossor] T_2$, but it is technically a bit more
complicated; we omit the details here. With these definitions,
Theorem~\ref{th:composition} holds for extended test sets in place
of adequate sets and for $\alpha = \crossor$ or $\alpha =
\unionor$. Thus we can extend the approach in
section~\ref{sec:decomposition} to large instances formed using $[
\crossor ]$ and $[ \unionor ]$. However, the operations $[
\crossor ]$ and $[ \unionor ]$ do not seem as useful in forming
combined instances; typically one wishes to use the conjunction,
not disjunction, of constraints.

\section{Implementation issues}
    \label{sec:implement}

We have built a prototype implementation of a tool for generating
adequate test sets.
The tool reads an instance $I$ of the test selection problem,
and produces a minimal adequate set for $I$.
The instances accepted by the tool are specified
in the concrete language of Section~\ref{subsec:enum};
the criteria \ea\ and \exh\ are also allowed, and are automatically
converted to expressions that use only \union\ and \cross.

Internally, the tool works in six phases:
\begin{enumerate}
   \item
      \label{phase:parse}
      Parse the input and check its consistency
      (only declared parameters and values are used,
      no parameter is declared twice, etc.).
   \item
      \label{phase:elim}
      Eliminate \ea\ and \exh.
   \item
      \label{phase:transcrit}
      Transform the criterion to the \union\cross\ form.
   \item
      \label{phase:transcons}
      Transform the constraint to the $\vee\wedge$ form.
   \item
      \label{phase:findmin}
      Find a minimal adequate set, using the algorithm in
         Figure~\ref{fig:algMal}.
   \item
      \label{phase:out}
      Print the test points.
\end{enumerate}

The tool is implemented in C;
the total size of the source files is about 1200 lines.
The basic data structures are trees and forests,
which are used to represent the parsed declarations, constraints
and criteria, as well as the intermediate results
for the transformations in phases~\ref{phase:transcrit}
and~\ref{phase:transcons}.

We have tested the tool on RISC System/6000 Model 560,
under the AIX operating system.\footnote{RISC System/6000 and AIX
are trademarks of International Business Machines Corporation.}
To measure the execution time on instances with large
minimal adequate test sets, we have used the criterion \exh.
For the instance in Figure~\ref{fig:test}, the domain $D(\constr)$
has 3125 points, and the only adequate set is the whole domain.
Although this is a very special form of a test selection criterion,
the tool does not take any shortcuts;
instances like this one are therefore suitable for performance
measurements.
\begin{figure}[tb]

\fuline
\begin{tabbing}
\hspace{8mm}\=\hspace{5mm}\=\hspace{8mm}\=\hspace{8mm}\=\hspace{3cm}\=
                                                               \kill
{\sf declaration}
\\[-1.5mm] \>{\bf Alice} \>\>: \{ {\sc a1, a2, a3, a4, a5} \,\}
\\[-2.5mm] \>{\bf Bob  } \>\>: \{ {\sc b1, b2, b3, b4, b5} \,\}
\\[-2.5mm] \>{\bf Cathy} \>\>: \{ {\sc c1, c2, c3, c4, c5} \,\}
\\[-2.5mm] \>{\bf Diana} \>\>: \{ {\sc d1, d2, d3, d4, d5} \,\}
\\[-2.5mm] \>{\bf Elaine} \>\>: \{ {\sc e1, e2, e3, e4, e5} \,\}
\\[-1.5mm] {\sf criterion}
\\[-1.5mm] \>\exh( {\bf Alice}, {\bf Bob}, {\bf Cathy}, {\bf
Diana}, {\bf Elaine} )
\end{tabbing}
\fuline
\caption{An instance to generate 3125 test points}
\label{fig:test}
\end{figure}
The execution time of the tool for this input is slightly less than 30 seconds
--- that is, more than 100 test points per second.
By using more sophisticated data structures
we would be able to improve this number substantially;
however, enhancing the functionality of the tool
is more important than optimizing its running time.

In particular, it would be worthwhile to extend the language
with other data types (see the discussion of future work
in Section~\ref{subsec:future}).
Other possible enhancements would be to add heuristics
to the test selection algorithm,
and to compute bounds on the size of the test set before
the selection algorithm is invoked.

Although the tool always produces a {\em minimal\/} adequate set,
it makes no attempt to come close to a {\em minimum\/} adequate set.
A more sophisticated implementation would include {\bf heuristics}
to make the generated set smaller in ``typical cases''.
A simple heuristic of this kind is to order the subcubes in the set
$S$ in Figure~\ref{fig:algMal}
so that smaller subcubes are processed before larger ones.

An approximate {\bf bound\/} for the size of the produced test set
would be useful as an early feedback to the user
when the tool is used on a large instance.
The user would appreciate some estimate of the size of the test set
before the test selection algorithm itself is run.
An upper bound can be easily computed as follows,
even before phase~\ref{phase:transcrit} begins:
In the criterion expression, replace each primitive criterion
by the value 1, replace each \union\ by the operator $+$,
and each \cross\ by the operator $\times$.
Then evaluate the resulting arithmetic expression;
the result is an upper bound for the size of the minimal test set
produced by the tool.
An enhanced version of the tool would first display
an initial (pessimistic) upper bound on the size of the test set,
and then update the bound as the computation progresses.
The designer could abandon execution if the bound seemed
hopelessly large.

\section{Concluding remarks}
    \label{sec:conclude}

\subsection{Related work}
    \label{subsec:related}

As is pointed out in the introduction,
the representation of test selection criteria by sets of subsets of
the input domain was considered, implicitly or explicitly,
by a number of researchers.
In {\em partition testing\/}~\cite{jeng},
the input domain is partitioned into subsets,
and one test point is then selected in each subset.
This is an elaboration of the {\em condition table method\/}
of Goodenough and Gerhart~\cite{A0894}.
In this line of research,
the emphasis has been on rules for constructing criteria from
program texts and specifications.
In contrast, the emphasis in the present paper is on a {\em language\/}
for specifying criteria (i.e. sets of subdomains),
and on operations that allow test designers to combine criteria.

In his discussion of functional testing,
Howden~\cite{howden} stresses the need to identify input domains,
and gives guidelines for systematic selection of test points
for several types of input values that occur in scientific programs.
Our basic philosophy is similar to Howden's;
we develop this point of view further, by automating part
of the selection process.

An important technical point is that we do not attempt to represent
a criterion by a set of {\em disjoint\/} subsets.
Note that our operation \union\ would make little sense
if we only considered sets of disjoint subsets.
As is explained by Jeng and Weyuker~\cite{jeng},
many naturally arising test selection criteria
lead to non-disjoint sets of subdomains.

Gourlay~\cite{gourlay} presents a precise framework for the discussion
of issues in testing.
In his terminology, our test selection criteria
are a special form of the test methods for the set-choice construction
testing system.
Gourlay reinterprets previously published discussions
about the suitability of various test selection criteria.
In our approach, we do not attempt to decide
a priori which criteria are sufficient
--- we leave that decision to the test designer.
That is why we emphasize the importance of a {\em language\/}
in which criteria are specified.

\subsection{Comparison with TSL}
    \label{subsec:compareTSL}

Balcer, Hasling and Ostrand~\cite{balcer} describe a complete test
language, called TSL, in which the test designer specifies a template for
the test cases to be generated, {\em categories\/}
(i.e. parameters and environment conditions),
choices of values for the categories,
and results of the test cases.
A TSL specification is automatically translated to a set of individual
test cases.

We now explain how TSL relates to the languages for test selection criteria
that we propose in this paper.
We will not describe TSL here; the reader is referred to the original
paper~\cite{balcer} for a detailed description.

A TSL specification contains declarations of parameters,
each with a set of values.
(TSL makes a distinction between parameters and environment conditions,
but for the purpose of this discussion both are considered to be
parameters.)
The specification also contains a set of Boolean conditions
(the IF clauses in the RESULT sections), which are used to decide
what combinations of parameter values are to be selected to
form test cases.
There are two types of such conditions:
unqualified ones, and
those qualified by the directive SINGLE.

Let us first consider the following simplified form of the
test selection criterion used by TSL:
For an unqualified condition, all combinations of parameter values satisfying
the condition should be selected.
For a qualified condition, at least one combination of parameter values
should be selected.
We show how to specify this criterion in our language.
Let $ \varphi_1 , \ldots, \varphi_m $ be the unqualified conditions,
and let $ \sigma_1 , \ldots, \sigma_r $ be the conditions qualified as SINGLE.
The test selection criterion is
\begin{equation}
    \fromto{6mm}{\bigunion}{i=1}{m} \;
      ( \;\langle \varphi_i \rangle \;\cross\; \exh\ \; )
    \;\; \union \;\;
    \fromto{6mm}{\bigunion}{j=1}{r} \;
       \langle \sigma_j \rangle
                           \label{eq:tslcr}
\end{equation}
If all $ \varphi_i $ and $ \sigma_j $ are conjunctions
of conditions of the form
\begin{eqnarray*}
    q & = & c    \\
    q & \neq & c
\end{eqnarray*}
where $q$ is a parameter and $c$ is a value of $q$,
then the criterion~(\ref{eq:tslcr}) can be expressed in the concrete
language from Section~\ref{subsec:enum}.

The TSL criterion as stated in~\cite{balcer} is actually more
complicated than the one in the previous paragraph.
An error-sensitizing rule is used to constrain the choice of a test
point for $ \langle \sigma_j \rangle $.
The rule is described only informally in~\cite{balcer};
we now state one possible formalization, using our language.
For each $ \sigma_j $, $j=1,\ldots,r$,
let $ \omega_j $ be the disjunction of all $ \varphi_i $ and
$ \sigma_i $ in the same RESULT section, other than $ \sigma_j $ itself.
The modified test selection criterion is
\[
    \fromto{6mm}{\bigunion}{i=1}{m} \;
      ( \;\langle \varphi_i \rangle \;\cross\; \exh\ \; )
    \;\; \union \;\;
    \fromto{6mm}{\bigunion}{j=1}{r} \;
       \langle \sigma_j \wedge \neg \, \omega_j \rangle
\]
It is not our goal to discuss the merits of various versions of the
error-sensitizing rule.
We merely make the point that our language is a convenient notation
for stating such rules precisely.

The language scheme proposed in this paper indicates the direction in which
the TSL notation for test selection, and other similar notations,
could be extended.
The test designer would benefit from the flexibility of
the operations \union\ and \cross.
For instance, in the example in Section~\ref{sec:example},
suppose that the test designer wants to fix
$ \sepa = \vslash $,
$ \sepb = \vslash $ and $ \aoccurs = \true $, and
test all values of \stra\ except \vempty\
and all values of \strb\ at least once,
but not necessarily all combinations of \stra\ and \strb.
The criterion to express that requirement is
\begin{eqnarray*}
 & &       \langle \sepa = \vslash \rangle
 \;\cross\; \langle \sepb = \vslash \rangle
 \;\cross\; \langle \aoccurs = \true \rangle     \\
 & \cross & \langle \stra \neq \vempty \rangle
 \;\cross\; \left( \;\exh( \stra ) \;\union\; \exh( \strb ) \; \right)
\end{eqnarray*}

\subsection{Future work}
    \label{subsec:future}

Here we mention several topics for further research
which we have not addressed in the present paper.
We group the topics into two categories:
Improved algorithms for the concrete language,
and extensions of the language and its use.

\noindent
$\rhd \;\;${\bf Algorithms for our concrete language}

In Section~\ref{sec:decomposition} we describe an algorithm for
discovering a decomposition into independent instances.
We assume that the constraint has the form $ \bigwedge_k \constr_{k} $.
To what extent can that assumption be relaxed?

Consider only the instances of the test selection problem that are
built from instances of some small bounded size using the
operations $ [ \crossand ] $ and $ [ \unionand ] $. Is there an
efficient algorithm for finding minimum adequate sets for the
instances in this special form?

Heuristics for finding ``almost-minimum'' adequate test sets
for ``common'' test selection criteria should be investigated.
In view of the results in Section~\ref{sec:color},
known heuristics for graph coloring would be a good starting point.

\noindent
$\rhd \;\;${\bf Extensions of the language}

The general language schema in Section~\ref{subsec:schema}
is a framework for further design of concrete languages
based on other data types.
After the enumerated data types treated in Section~\ref{subsec:enum},
the next most important type is {\bf integers\/}.
Some useful criteria for integers were mentioned in~\cite{notation},
but we have not studied in detail the algorithms needed to deal
with those criteria.

Another important candidate for incorporation into the general
schema is the type {\bf words over a finite alphabet\/},
which would be useful for specifying criteria that have to do
with control flow in a program or in a state machine.

The ideas in Section~\ref{sec:decomposition} lead naturally to
{\em modular\/} descriptions of complex test suites.
In a testing system supporting modularity, parameterized test
cases along with test selection criteria could be created
for various subsystems of a complex implementation under test,
independently of each other
(perhaps written by different test designers),
and then combined by means of simple operators.


\begin{thebibliography}{99}

\bibitem{AHU}
{\sc A.V.~Aho, J.E.~Hopcroft and J.D.~Ullman.}
{\em The Design and Analysis of Computer Algorithms.}
Addison-Wesley Publishing Co.~1974.

\bibitem{balcer}
{\sc M.J.~Balcer, W.M.~Hasling and T.J.~Ostrand.}
Automatic generation of test scripts from formal test specifications.
{\em Proc. Third ACM SIGSOFT Symp. Software Testing, Analysis and
Verification\/}
(Key West, Florida, December 13-15, 1989),
210-218.

\bibitem{ceg}
{\sc W.R.~Elmendorf.}
Cause-effect graphs in functional testing.
{\em IBM Poughkeepsie Laboratory, Technical Report TR-00.2487\/}
(November 1973).

\bibitem{GJ}
{\sc M.R.~Garey and D.S.~Johnson.}
{\em Computers and Intractability: A Guide to the Theory of
NP-Completeness.}
W.H.~Freeman and Co., New York 1979.

\bibitem{A0894}
{\sc J.B.~Goodenough and S.L.~Gerhart.}
Toward a theory of test data selection.
{\em IEEE Trans. Software Engineering SE-1}
(1975), 156-173.

\bibitem{gourlay}
{\sc J.S.~Gourlay.}
A mathematical framework for the investigation of testing.
{\em IEEE Trans. Software Engineering SE-9}
(1983), 686-709.

\bibitem{hamlet}
{\sc R.~Hamlet.}
Theoretical comparison of testing methods.
{\em Proc. Third ACM SIGSOFT Symp. Software Testing, Analysis and
Verification\/}
(Key West, Florida, December 13-15, 1989),
28-37.

\bibitem{howden}
{\sc W.E.~Howden.}
Functional program testing.
{\em IEEE Trans. Software Engineering SE-6}
(1980), 162-169.

\bibitem{jeng}
{\sc B.~Jeng and E.J.~Weyuker.}
Some observations on partition testing.
{\em Proc. Third ACM SIGSOFT Symp. Software Testing, Analysis and
Verification\/}
(Key West, Florida, December 13-15, 1989),
38-47.

\bibitem{myers}
{\sc G.J.~Myers.}
{\em The Art of Software Testing.}
John Wiley \& Sons, Inc.~1979.

\bibitem{notation}
{\sc J.~Pachl.}
A notation for specifying test selection criteria.
{\em Protocol Specification, Testing and Verification X\/}
(L.~Logrippo, R.L.~Probert and H.~Ural, editors; North-Holland 1990),
71-84.

\bibitem{philips}
{\sc T.K.~Philips.}
New algorithms to color graphs and find maximum cliques.
{\em IBM Research Division, Research Report RC~16326}
(November 1990).

\bibitem{toft}
{\sc B.~Toft.} Colouring, stable sets and perfect graphs. {\em
Handbook of Combinatorics\/}, to appear.

\end{thebibliography}
\end{document}